\documentclass[twocolumn,showpacs,amsmath,amssymb,pra,nofootinbib]{revtex4}
\usepackage{theorem}
\usepackage{graphicx}
\usepackage{color}

\theorembodyfont{\rmfamily}   
\newtheorem{theorem}{Theorem}

\newtheorem{definition}{Definition}

\newtheorem{proposition}{Proposition}

\newcommand{\qed}{\hfill$\square$}

\newenvironment{proof}{%
  \noindent{\em Proof.\ }}{%
  \hspace*{\fill}\qed \\
  \vspace{2ex}}
\DeclareMathAlphabet{\bm}{OML}{cmm}{b}{it}
\newcommand{\ket}[1]{| #1 \rangle} 
\newcommand{\bra}[1]{\langle #1 |}

\newcommand{\rom}[1]{\mathrm{#1}}

\begin{document}

\title{Private and Quantum Capacities of More Capable and Less Noisy Quantum Channels}

\author{Shun Watanabe}
 \email{shun-wata@is.tokushima-u.ac.jp}
 \affiliation{%
Department of Information Science and Intelligent Systems, 
University of Tokushima, \\
2-1, Minami-josanjima, Tokushima, 770-8506 Japan
}

\date{October 26, 2011}

\begin{abstract}
Two new classes of quantum channels, which we call {\em more capable} 
 and {\em less noisy}, are introduced. The more capable class consists
of channels such that the quantum capacities of the complementary channels
to the environments are zero. The less noisy class consists
of channels such that the private capacities of the complementary
channels to the environment are zero.
For the more capable class, it is clarified that the private
capacity and quantum capacity coincide. For the less noisy class,
it is clarified that the private capacity and quantum capacity 
can be single letter characterized.
\end{abstract}

\maketitle

\section{Introduction}         

One of the most important problem in 
quantum information theory is to determine 
the quantum capacity of a noisy quantum channel.
The capacity is defined as the transmission rate  optimized over all possible
quantum error correcting codes such that decoding
errors vanish in the limit of asymptotically many uses of the channel.

Mathematically, a quantum channel can be
described by the trace preserving completely positive (TPCP) map from the input system to
the output system. By using the Stinespring dilation of the
TPCP map, we can naturally define a complementary channel
to an environment system, and we can regard the noisy quantum channel as 
a wire-tap channel \cite{wyner:75,csiszar:78} from the sender to the legitimate receiver 
and the eavesdropper who can observe the environment system of the channel
(eg.~see \cite{hayashi-book:06}). Then we can define the private
capacity of the noisy quantum channel as the transmission rate optimized over all
possible wire-tap codes such that decoding errors and
information leakage vanish in the limit of asymptotically
many uses of the channel.

The private capacity and quantum capacity of noisy quantum
channels were established in \cite{lloyd:97,shor:02,devetak:05,cai:04}.
However unlike the capacity formula of a classical noisy channel
or the private capacity formula of a classical wire-tap channel,
the private capacity and quantum capacity formulae are not
single letter characterized, i.e., the formulae involve 
the limit with respect to the number of channel uses, and they
are not computable. Indeed, some numerical evidences clarified that 
the expressions in the capacity formulae are not additive
\cite{shor:96,divincenzo:98,smith:07b,smith:08}, and the single
letter characterization is not possible in general at least
by using the same expressions.

A quantum channel is called degradable if there exists
another degrading channel such that the conjunction of
the channel to the legitimate receiver and the degrading channel 
coincide with the complementary channel to the eavesdropper.
In such a case, the single letter characterizations of
the private capacity and quantum capacity were 
established \cite{devetak:05b,hayashi-book:06}.

A quantum channel is called conjugate degradable if
there exists another degrading channel such that 
the conjunction of the channel to the legitimate receiver
and the degrading channel coincide with the complementary channel to the 
eavesdropper up to complex conjugation. In such a case, 
the single letter characterizations were also established \cite{bradler:10}.

To date, all quantum channel whose capacities are single letter 
characterized are degradable or
conjugate degradable\footnote{There are also channels called
anti-degradable or conjugate anti-degradable. The capacities of those channels are also
single letter characterized, but the capacities are equal to zero.},
and it is important to clarify a broader class of quantum channels such that
the single letter characterizations are possible.

Aside from the possibility of the single letter characterizations,
there is also another interesting problem.
In the quantum information theory, the private information transmission
and the quantum information transmission are closely 
related \cite{schumacher:96,schumacher:98,shor:00,devetak:05}, and the possibility
of the latter implies the possibility of the former.
However, the private information transmission and the quantum
information transmission are not exactly equivalent. 
Indeed, although the private capacity and quantum capacity coincide for
degradable quantum channels \cite{smith:07},
the former can be strictly larger than 
the latter in general. Especially the private capacity
can be positive even if the quantum capacity is zero \cite{horodecki:08}.
Thus it is important to clarify a condition on quantum channels
such that the private capacity and quantum capacity coincide or not.

To shed light on the above mentioned
two problems, we introduce two new classes of 
quantum channels, which we call {\em more capable}
and {\em less noisy}. The more capable class consists of
channels such that the quantum capacities of the
complementary channels are zero. The less noisy class
consists of channels such that the private capacities of the
complementary channels are zero. 
Later, these definitions turn out to be natural
analogies of the partial orderings, more capable
and less noisy, between classical channels \cite{korner:75}.

\begin{figure}
\centering
\includegraphics[width=0.5\linewidth]{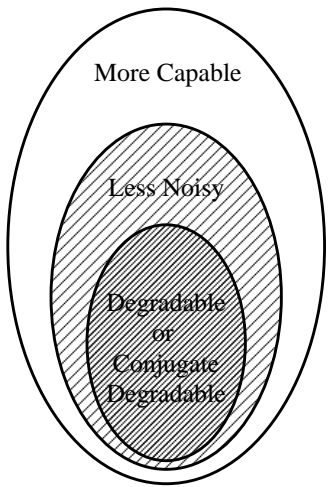}
\caption{The inclusive relation of the degradable, the conjugate degradable,
the less noisy, and the more capable classes of quantum channels.}
\label{Fig:quantum-channel}
\end{figure}

The inclusive relation of the degradable, the conjugate
degradable, the less noisy, and the more capable
classes are summarized in Fig.~\ref{Fig:quantum-channel}.
In this paper, we show that the private capacity and 
quantum capacity coincide for channels in the more capable class.
Furthermore, we also show that the private capacity and 
quantum capacity can be single letter characterized 
for channels in the less noisy class.
These results provide partial solutions to the above
mentioned two problems.

The rest of the paper is organized as follows.
In Section \ref{section:preliminaries}, we review some known results on
the private capacity and quantum capacity of 
quantum channels. In Section \ref{section:main}, 
the more capable and less noisy 
classes are introduced, and we state our main results.
In Section \ref{section:proof}, we summarize 
certain properties implied by more capable and less noisy,
and show proofs of our main results.
We finalize the paper with conclusion in Section \ref{section:proof}.

\section{Preliminaries}
\label{section:preliminaries}

Let ${\cal N}_B$ be a quantum channel from an input system ${\cal H}_A$ to
an output system ${\cal H}_B$. 
By using the Stinespring dilation (eg.~see \cite{hayashi-book:06}),
there exist an environment system ${\cal H}_E$ and an isometry
$U_{BE}$ from ${\cal H}_A$ to the joint system 
${\cal H}_B \otimes {\cal H}_E$ such that
\begin{eqnarray*}
{\cal N}_B(\rho) = \rom{Tr}_E\left[ U_{BE} \rho U_{BE}^* \right]
\end{eqnarray*}
for every input $\rho$, where $\rom{Tr}_E$ is the partial trace with
respect to the environment system. By using this representation, we 
can naturally define another channel
\begin{eqnarray*}
{\cal N}_E(\rho) = \rom{Tr}_B\left[ U_{BE} \rho U_{BE}^* \right],
\end{eqnarray*}
which is usually called the complementary channel of ${\cal N}_B$.
Although the Stinespring dilation is not unique, the following
arguments do not depend on the choice of the dilation because
two dilations can be converted to each other by applying a local
unitary to the environment systems.

Throughout the paper, we basically follow the notations from
\cite{nielsen-chuang:00,hayashi-book:06}.
The von Neumann entropy of a density matrix $\rho$ is defined by
$H(\rho) = - \rom{Tr} \rho \log \rho$, and the quantum
relative entropy between $\rho$ and $\sigma$ is defined by
$D(\rho \| \sigma) = \rom{Tr} \rho(\log \rho - \log \sigma)$.
For input state $\rho_A$ to the channel ${\cal N}_B$, the 
coherent information is defined by 
$I_c(A \rangle B)_\rho = H({\cal N}_B(\rho_A)) - H({\cal N}_E(\rho_A))$.
When the input state is clear from the context, we omit the subscript
and denote $I_c(A \rangle B)$.
The quantum mutual information of $\rho_{XB}$ on the joint system is defined
by $I(X; B) = H(\rho_X) + H(\rho_B) - H(\rho_{XB})$.
Especially, when $\rho_{XB}$ is classical with respect to $X$, i.e.,
$\rho_{XB}$ is of the form
\begin{eqnarray*}
\rho_{XB} = \sum_x P_X(x) \ket{x}\bra{x} \otimes \rho_B^x,
\end{eqnarray*}
then the quantum mutual information can be written as 
\begin{eqnarray*}
I(X;B) = H(\rho_B) - \sum_x P_X(x) H(\rho_B^x).
\end{eqnarray*}

When the legitimate receiver can observe the output of ${\cal N}_B$
and the eavesdropper can observe the output of ${\cal N}_E$, 
the private capacity is characterized by \cite{cai:04,devetak:05}
\begin{eqnarray}
\label{eq:private-capacity-formula}
	C_p({\cal N}_B) = \lim_{n \to \infty} \frac{1}{n} C_p^{(1)}({\cal N}_B^{\otimes n}),
\end{eqnarray}
where
\begin{eqnarray*}
	C_p^{(1)}({\cal N}_B) := \max_{P_{U}, \{\rho_A^u\}} [ I(U; B) - I(U;E)],
\end{eqnarray*}
where $\{ \rho_A^u \}$ are states (not necessarily pure states)
on ${\cal H}_A$ indexed by $u \in 
{\cal U}$, and $P_U$ is a probability distribution on a finite set ${\cal U}$.

On the other hand, when the sender want to transmit quantum information
to the receiver through the channel ${\cal N}_B$, the quantum 
capacity is characterized by \cite{lloyd:97,devetak:05}
\begin{eqnarray}
\label{eq:quantum-capacity-formula}
	Q({\cal N}_B) = \lim_{n \to \infty} \frac{1}{n} Q^{(1)}({\cal N}_B^{\otimes n}),
\end{eqnarray}
where 
\begin{eqnarray*}
	Q^{(1)}({\cal N}_B) := \max_{ \rho_A } I_c(A \rangle B)_\rho.
\end{eqnarray*}

Both Eqs.~(\ref{eq:private-capacity-formula}) 
and (\ref{eq:quantum-capacity-formula}) cannot be single letter characterized,
i.e., $C_p({\cal N}_B) > C_p^{(1)}({\cal N})$ and 
$Q({\cal N}) > Q^{(1)}({\cal N})$
in general \cite{shor:96,divincenzo:98,smith:07b,smith:08}.
Furthermore, the private capacity can be
strictly larger than the quantum capacity, i.e.,
$C_p({\cal N}_B) > Q({\cal N}_B)$ for some channels \cite{horodecki:08}.

The channel ${\cal N}_B$ is said to be degradable if there exists a TPCP 
map ${\cal D}$ such that
${\cal D} \circ {\cal N}_B = {\cal N}_E$. 
This is a quantum analogue of degraded broadcast channel \cite{cover}.
When ${\cal N}_B$ is degradable, then it is known \cite{devetak:05b,hayashi-book:06} 
that the single letter formulae hold, i.e., $C_p({\cal N}_B) = C_p^{(1)}({\cal N}_B)$ and 
$Q({\cal N}_B) = Q^{(1)}({\cal N}_B)$. Furthermore, it is also known that
$C_p({\cal N}_B) = Q({\cal N}_B)$ for degradable channel ${\cal N}_B$ \cite{smith:07}.

Let ${\cal C}$ denote entry-wise complex conjugation with respect to
a fixed basis of ${\cal H}_E$. Then, the channel ${\cal N}_B$ is said to be
conjugate degradable if there exists a TPCP map ${\cal D}$ such that
${\cal D} \circ {\cal N}_B = {\cal C} \circ {\cal N}_E$.
When ${\cal N}_B$ is conjugate degradable, it is known that
$Q({\cal N}_B) = Q^{(1)}({\cal N}_B)$ \cite{bradler:10}.
Later, it will turn out  that
$C_p({\cal N}_B) = Q({\cal N}_B) = Q^{(1)}({\cal N}_B)$ also holds.

\section{Main Statements}
\label{section:main}

In this section, we introduce two new classes of
quantum channels, and show our main results.
\begin{definition}
The quantum channel ${\cal N}_B$ is said to be 
{\em more capable} than the environment, or just
more capable, if the quantum capacity
of the complementary channel to the environment is zero, i.e.,
\begin{eqnarray}
\label{eq:q-more-capable}
Q({\cal N}_E) = 0.
\end{eqnarray}
\end{definition}
\begin{definition}
The quantum channel ${\cal N}_B$ is said to be 
{\em less noisy} than the environment, or just
less noisy, if the private capacity
of the complementary channel to the environment is zero, i.e.,
\begin{eqnarray}
\label{eq:q-less-noisy}
C_p({\cal N}_E) = 0.
\end{eqnarray}
\end{definition}
Since $C_p({\cal N}_E) \ge Q({\cal N}_E)$, 
less noisy implies more capable.

By using the eigenvalue decomposition 
\begin{eqnarray*}
\rho_{A^n} = \sum_{x^n \in {\cal X}^n} P_{X^n}(x^n) \ket{\psi_{x^n}}\bra{\psi_{x^n}}
\end{eqnarray*}
of $\rho_{A^n}$, we can rewrite the coherent information as
\begin{eqnarray}
I_c(A^n \rangle B^n) = I(X^n; B^n) - I(X^n; E^n),
\label{eq:alternative-coherent}
\end{eqnarray}
where we set $|{\cal X}| = \dim {\cal H}_A$. Thus, by noting Eq.~(\ref{eq:quantum-capacity-formula}),
the quantum channel 
${\cal N}_B$ being more capable can be also described as 
\begin{eqnarray}
\label{eq:q-more-capable-alternative}
I(X^n; B^n) \ge  I(X^n; E^n),~~\forall (P_{X^n}, \{ \ket{\psi_{x^n}} \})
\end{eqnarray}
holds for every $n \ge 1$.
Furthermore, by noting Eq.~(\ref{eq:private-capacity-formula}),
the quantum channel 
${\cal N}_B$ being less noisy can be also described as 
\begin{eqnarray}
\label{eq:q-less-noisy-alternative}
	I(U^n;B^n) \ge I(U^n; E^n),~~\forall (P_{U^n},\{\rho_{A^n}^{u^n} \})
\end{eqnarray}
holds for every $n \ge 1$.
Eqs.~(\ref{eq:q-more-capable-alternative}) and (\ref{eq:q-less-noisy-alternative})
resemble the definitions of more capable and less noisy for 
classical channels \cite{korner:75}, and it is justified to call 
quantum channels satisfying Eqs.~(\ref{eq:q-more-capable}) or (\ref{eq:q-less-noisy})
more capable or less noisy. 

In \cite{korner:75}, alternative description of less noisy,
less divergence contracting, was introduced, and we can 
also extend such a description to the 
quantum channel. The quantum channel ${\cal N}_B$ is said to be
less divergence contracting if 
\begin{eqnarray}
\lefteqn{
D({\cal N}_B^{\otimes n}(\rho_{A^n}) \| {\cal N}_B^{\otimes n}(\hat{\rho}_{A^n})) } 
\nonumber \\
 &\ge& D({\cal N}_E^{\otimes n}(\rho_{A^n}) \| {\cal N}_E^{\otimes n}(\hat{\rho}_{A^n})),~~~
\forall \rho_{A^n}, \hat{\rho}_{A^n}
\label{eq:q-less-divergence-contracting}
\end{eqnarray}
holds for every $n \ge 1$. 
Later, we will show that the quantum channel is 
less noisy if and only if less divergence contracting
(Proposition \ref{proposition:q-equivalence-less-noisy-divergence}).
This alternative description plays a crucial role when
we prove Theorem \ref{theorem:less-noisy}.

The followings are our main results.
\begin{theorem}
\label{theorem:more-capable}
Suppose that the quantum channel ${\cal N}_B$ is more
capable. Then, we have
\begin{eqnarray*}
C_p({\cal N}_B) = Q({\cal N}_B).
\end{eqnarray*}
\end{theorem}
\begin{theorem}
\label{theorem:less-noisy}
Suppose that the quantum channel ${\cal N}_B$ is less noisy.
Then, we have
\begin{eqnarray*}
C_p({\cal N}_B) = Q({\cal N}_B) = Q^{(1)}({\cal N}_B).
\end{eqnarray*}
\end{theorem}

When ${\cal N}_B$ is conjugate degradable,
we can show that $C_p({\cal N}_E) = 0$ as follows.
Suppose that the sender sends a state $\rho_i$
that corresponds to the message $i$, and 
the eavesdropper\footnote{The role of the legitimate receiver
and the eavesdropper is interchanged because we are considering
the private capacity of ${\cal N}_E$.} uses a POVM $\{ M_i \}$.
Then, for the entry-wise complex conjugate POVM $\{ \bar{M}_i \}$,
we have
\begin{eqnarray*}
\rom{Tr}[\bar{M}_i {\cal D}^{\otimes n} \circ {\cal N}_B^{\otimes n}(\rho_i)]
 &=& \rom{Tr}[\bar{M}_i {\cal C}^{\otimes n} \circ {\cal N}_E^{\otimes n}(\rho_i)] \\
 &=& \rom{Tr}[M_i {\cal N}_E^{\otimes n}(\rho_i)],
\end{eqnarray*}
where the last equality follows because 
$\bar{M}_i^T = M_i$ and $({\cal C}^{\otimes n} \circ {\cal N}_E^{\otimes 
n}(\rho_i))^T = {\cal N}_E^{\otimes n}(\rho_i)$.
Thus, the legitimate receiver can get exactly the same information
as the eavesdropper and the private information transmission to
the eavesdropper is impossible.
From this argument, conjugate degradable 
implies less noisy.

When the quantum capacity of the channel is $0$ but
it can be used to share bound entanglement, then the
channel is called a binding entanlement channel \cite{horodecki:00}.
Especially when the channel produces a 
positive partial transpose (PPT) bound entanglement, the channel
is called PPT binding entanglement channel. 
If a complementary channel is a binding entanglement channel,
then the main channel obviously belongs to the more capable class.
Since the complementary channel of the conjugate degradable channel can
only produce PPT bipartite state \cite{bradler:10}, if there exists a
conjectured negative partial transpose (NPT) binding entanglement channel,
the complementary of such a channel belongs to the more capable class but
not to the conjugate degradable class.

It is known that there exists a channel such 
that the quantum capacity is zero (PPT binding entanglement channel) but
the private capacity is strictly positive \cite{horodecki:08}.
Let the complementary channel ${\cal N}_E$ be such a channel.
Then the channel ${\cal N}_B$ is more capable but not less 
noisy\footnote{Note that the private and quantum capacities of
this channel is strictly positive, which can be checked as follows.
If $Q({\cal N}_B) = 0$,
then the complementary channel ${\cal N}_E$ is more capable.
Then, Theorem \ref{theorem:more-capable} implies that
$Q({\cal N}_E) = C_p({\cal N}_E)$, which contradict with the fact
$C_p({\cal N}_E) > Q({\cal N}_E) = 0$.}. 
Thus, the more capable class is strictly broader than
the less noisy class. 
However, it is not yet clear whether the less noisy class is
strictly broader than the degradable or conjugate degradable classes.

\section{Proof of Theorems}
\label{section:proof}

\subsection{Properties of $C_p^{(1)}({\cal N}_B)$ and $Q^{(1)}({\cal N}_B)$}

In this section, we summarize the properties of
$C_p^{(1)}({\cal N}_B)$ and $Q^{(1)}({\cal N}_B)$
when Eqs.~(\ref{eq:q-more-capable-alternative}) 
or (\ref{eq:q-less-noisy-alternative}) hold for $n=1$.
For $n \ge 2$, we can also show similar properties of $C_p^{(1)}({\cal N}_B^{\otimes n})$ 
and $Q^{(1)}({\cal N}_B^{\otimes n})$ 
when Eqs.~(\ref{eq:q-more-capable-alternative}) 
or (\ref{eq:q-less-noisy-alternative}) hold for
each $n$ by considering $n$ times extension of ${\cal N}_B$.
The following properties can be regarded as
quantum extensions of the properties shown for classical channels in
the literatures \cite{korner:75,csiszar:78,dijk:97,ozel:11}

\begin{proposition}
\label{proposition:q-more-capable-no-preprocessing}
Suppose that Eq.~(\ref{eq:q-more-capable-alternative}) holds for $n=1$. 
Then we have
\begin{eqnarray*}
C_p^{(1)}({\cal N}_B) = Q^{(1)}({\cal N}_B).
\end{eqnarray*}
\end{proposition}
\begin{proof}
For any $P_U$ and $\{ \rho_A^u \}$, let 
\begin{eqnarray*}
\rho_A^u = \sum_x \alpha_{u,x} \ket{\psi_{u,x}}\bra{\psi_{u,x}}
\end{eqnarray*}
be an eigenvalue decomposition. Let $\tilde{X}$ be the random
variable on ${\cal U} \times {\cal X}$ such that
\begin{eqnarray*}
P_{\tilde{X}|U}(u^\prime,x|u) = \left\{
\begin{array}{ll}
\alpha_{u,x} & \mbox{if } u = u^\prime \\
0 & \mbox{else}
\end{array}
\right..
\end{eqnarray*}
Then, we have
\begin{eqnarray}
\lefteqn{
I(U;B) - I(U;E) = [ I(\tilde{X};B) - I(\tilde{X} ; E)] 
} \nonumber \\
&& - [ I(\tilde{X};B|U) - I(\tilde{X} ; E|U)].
\label{eq:q-more-capable-no-preprocessing-proof-1}
\end{eqnarray}
Since Eq.~(\ref{eq:q-more-capable-alternative}) holds for $n=1$, we have
\begin{eqnarray*}
I(\tilde{X} ; B|U=u) - I(\tilde{X} ; E|U=u) \ge 0
\end{eqnarray*}
for every $u$, which means
that the second bracket of Eq.~(\ref{eq:q-more-capable-no-preprocessing-proof-1})
is non-negative. Furthermore, by noting that 
$\{ \ket{\psi_{u,x}}\}$ are pure, we have 
\begin{eqnarray*}
I(\tilde{X};B) - I(\tilde{X} ; E) = I_c(A \rangle B),
\end{eqnarray*}
where 
\begin{eqnarray*}
\rho_A = \sum_{u,x} P_{U}(u) P_{\tilde{X}|U}(u,x|u) \ket{\psi_{u,x}}\bra{\psi_{u,x}}.
\end{eqnarray*}
Since $P_U$ and $\{ \rho_A^u\}$ are arbitrary, we have
\begin{eqnarray*}
C_p^{(1)}({\cal N}_B) \le Q^{(1)}({\cal N}_B).
\end{eqnarray*}
The opposite inequality is obvious from the definitions of
$C_p^{(1)}({\cal N}_B)$, $Q^{(1)}({\cal N}_B)$, and
Eq.~(\ref{eq:alternative-coherent}).
\end{proof}

\begin{proposition}
\label{proposition:q-channel-preprocessing}
Suppose that Eq.~(\ref{eq:q-more-capable-alternative}) does not hold for $n=1$, 
and the
density operator $\rho_{A}^*$ maximizing $I_c(A \rangle B)$
is full rank. Then, we have
\begin{eqnarray*}
C_p^{(1)}({\cal N}_B) > Q^{(1)}({\cal N}_B).
\end{eqnarray*}
Especially when $\dim {\cal H}_A = 2$ and $C_p^{(1)}({\cal N}_B) > 0$, the sufficient and required condition on
\begin{eqnarray*}
C_p^{(1)}({\cal N}_B) = Q^{(1)}({\cal N}_B)
\end{eqnarray*}
is that Eq.~(\ref{eq:q-more-capable-alternative}) holds for $n=1$.
\end{proposition}
\begin{proof}
Since Eq.~(\ref{eq:q-more-capable-alternative}) does not hold for $n=1$, there exists
$\hat{\rho}_A$ such that
\begin{eqnarray*}
I_c(A \rangle B)_{\hat{\rho}} < 0.
\end{eqnarray*} 
Since $\rho_A^*$ is full rank, there exists $0 < \lambda < 1$ such that
$\rho_A^* - \lambda \hat{\rho}_A$ is positive semidefinite.
We construct $P_U$ and $\{ \rho_A^u\}$ as follows.
Let 
\begin{eqnarray*}
\hat{\rho}_A &=& \sum_x \hat{P}_X(x) \ket{\hat{\psi}_x}\bra{\hat{\psi}_x}, \\
\rho_A^* - \lambda \hat{\rho}_A &=& \sum_x \beta_x \ket{\phi_x}\bra{\phi_x}
\end{eqnarray*}
be eigenvalue decompositions. Let ${\cal U} = \{0\} \cup {\cal X}$.
Then, we set $P_U(0) = \lambda$, $P_U(u) = \beta_u$ for $u \in {\cal X}$,
$\rho_A^0 = \hat{\rho}_A$, and 
$\rho_A^u = \ket{\phi_u}\bra{\phi_u}$ for $u \in {\cal X}$.
Let $\tilde{X}$ be the random variable on ${\cal U} \times {\cal X}$ 
 such that
\begin{eqnarray*}
P_{\tilde{X}|U}(u^\prime,x|u) = \left\{
\begin{array}{ll}
\hat{P}_X(x) & \mbox{if } u = u^\prime = 0 \\
1 & \mbox{if } x = u= u^\prime \neq 0 \\
0 & \mbox{else}
\end{array}
\right..
\end{eqnarray*}
Then, we have
\begin{eqnarray}
\lefteqn{
I(U;B) - I(U;E) 
} \nonumber \\
&=& I(\tilde{X};B) - I(\tilde{X};E) - [ I(\tilde{X};B|U) - 
 I(\tilde{X};E|U)] \nonumber \\
&=& I(\tilde{X};B) - I(\tilde{X};E) \nonumber \\
&& ~~~ - \lambda [I(\tilde{X};B|U=0) - I(\tilde{X};E|U=0)] 
  \nonumber \\
&>& I(\tilde{X};B) - I(\tilde{X};E) 
 \label{eq:proof-q-strict-preprocessing-1} \\
&=& I_c(A \rangle B)_{\rho^*}, 
 \label{eq:proof-q-strict-preprocessing-2}
\end{eqnarray}
where Eq.~(\ref{eq:proof-q-strict-preprocessing-1}) follows from
\begin{eqnarray*}
I(\tilde{X};B|U=0) - I(\tilde{X};E|U=0) = I_c(A \rangle B)_{\hat{\rho}} 
 < 0
\end{eqnarray*}
and Eq.~(\ref{eq:proof-q-strict-preprocessing-2}) follows from
\begin{eqnarray*}
\lefteqn{ \sum_x P_U(0) P_{\tilde{X}|U}(0,x|0) 
 \ket{\hat{\psi}_x}\bra{\hat{\psi}_x} } \\
 && + \sum_{u,x \in {\cal X}} P_U(u) P_{\tilde{X}|U}(u,x|u) 
 \ket{\phi_x}\bra{\phi_x} = \rho_A^*.
\end{eqnarray*}

Next, we show the latter statement of the proposition.
The sufficient condition follows from Proposition \ref{proposition:q-more-capable-no-preprocessing}.
Suppose that 
\begin{eqnarray}
\label{eq:proof-q-strict-improvement-1}
I_c(A \rangle B) \le 0,~~~\forall \rho_A
\end{eqnarray}
holds. Since $C_p^{(1)}({\cal N}_B) > 0$, there exists $P_U$ and
$\{ \rho_A^u\}$ such that 
\begin{eqnarray*}
I(U;B) - I(U;E) > 0,
\end{eqnarray*}
which implies the required condition. Next, we consider
the case such that Eq.~(\ref{eq:proof-q-strict-improvement-1})
does not hold. In this case, we have
\begin{eqnarray*}
\max_{\rho_A} I_c(A \rangle B) > 0.
\end{eqnarray*}
Then, since $\dim {\cal H}_A = 2$, $\rho_A^*$ must be full rank.
Thus, the required condition follows from the former statement
of the proposition.
\end{proof}

\begin{proposition}
\label{proposition:concavity}
Eq.~(\ref{eq:q-less-noisy-alternative}) holds for $n=1$ 
if and only if
the coherent information is concave\footnote{It should be noted that the 
 coherent information is known to be concave if the quantum channel 
 ${\cal N}_B$ is degradable \cite[Eq.~(9.89)]{hayashi-book:06}.}, i.e.,
\begin{eqnarray*}
I_c(A \rangle B)_{\rho} \ge \sum_{i=1}^m p_i I_c(A \rangle B)_{\rho_i},
\end{eqnarray*}
where $\rho = \sum_{i=1}^m p_i \rho_i$.
\end{proposition}
\begin{proof}
Let 
\begin{eqnarray*}
\rho_i = \sum_x p_{i,x} \ket{\psi_{i,x}}\bra{\psi_{i,x}}
\end{eqnarray*}
be an eigenvalue decomposition. Then, let ${\cal U} = \{1,\ldots,m\}$,
$P_U(u) = p_i$, $\tilde{X}$ be the random variable on ${\cal U} \times 
 {\cal X}$ such that
\begin{eqnarray*}
P_{\tilde{X}|U}(u^\prime,x|u) = \left\{
\begin{array}{ll}
p_{i,x} & \mbox{if } u^\prime = u \\
0 & \mbox{else}
\end{array}
\right..
\end{eqnarray*}
Then, we have
\begin{eqnarray}
\lefteqn{
I(U;B) - I(U;E) = [I(\tilde{X};B) - I(\tilde{X} ; E)] 
} \nonumber \\
&& - [I(\tilde{X} ; B|U) - I(\tilde{X} ; E|U)].
\label{eq:q-less-noisy-concavity-proof}
\end{eqnarray}
We also have
\begin{eqnarray*}
I(\tilde{X};B) - I(\tilde{X} ; E) = I_c(A\rangle B)_\rho
\end{eqnarray*}
and
\begin{eqnarray*}
I(\tilde{X};B|U) - I(\tilde{X};E|U) = \sum_{i=1}^m p_i I_c(A \rangle B)_{\rho_i}.
\end{eqnarray*}
Thus, from Eq.~(\ref{eq:q-less-noisy-concavity-proof}),
Eq.~(\ref{eq:q-less-noisy-alternative}) holds for $n=1$
if and only if the coherent information $I_c(A\rangle B)$ is
concave.
\end{proof}


\begin{proposition}
\label{proposition:q-equivalence-less-noisy-divergence}
The following two conditions are equivalent.
\begin{enumerate}
\renewcommand{\theenumi}{\roman{enumi}}
\renewcommand{\labelenumi}{(\theenumi)}
\item \label{q-condition1} Eq.~(\ref{eq:q-less-noisy-alternative}) holds for $n=1$.

\item \label{q-condition2} Eq.~(\ref{eq:q-less-divergence-contracting}) holds for $n=1$.
\end{enumerate}
\end{proposition}
\begin{proof}
We first show that (\ref{q-condition1}) implies (\ref{q-condition2}).
For any $\rho_A$, $\hat{\rho}_A$, and $0 \le \lambda \le 1$, let 
${\cal U} = \{0,1\}$, $P_{U_\lambda}(0) = \lambda$, $P_{U_\lambda}(1) = 1- \lambda$,
$\rho_A^0 = \rho_A$, and $\rho_A^1 = \hat{\rho}_A$. Then, let
\begin{eqnarray*}
\lefteqn{
f(\lambda) = I(U_\lambda;B) - I(U_\lambda;E)
} \\
&=& \lambda D({\cal N}_B(\rho_A) \| {\cal N}_B(\bar{\rho}_A))
  + (1-\lambda) D({\cal N}_B(\hat{\rho}_A)\| {\cal N}_B(\bar{\rho}_A)) \\
&-&  \lambda D({\cal N}_E(\rho_A)\| {\cal N}_E(\bar{\rho}_A))
  - (1-\lambda) D({\cal N}_E(\hat{\rho}_A)\| {\cal N}_E(\bar{\rho}_A)),
\end{eqnarray*}
where 
\begin{eqnarray*}
\bar{\rho}_A = \lambda \rho_A + (1-\lambda) \hat{\rho}_A.
\end{eqnarray*}
By elementary calculation (cf.~\cite[Ex.~1.4]{hayashi-book:06}), we have
\begin{eqnarray*}
f^\prime(0) = D({\cal N}_B(\rho_A)\| {\cal N}_B(\hat{\rho}_A))
  - D({\cal N}_E(\rho_A)\| {\cal N}_E(\hat{\rho}_A)).
\end{eqnarray*}
Obviously, we have $f(0) = 0$. 
Since 
Eq.~(\ref{eq:q-less-noisy-alternative}) holds for $n=1$, 
$f(\lambda) \ge 0$ for any $0\le \lambda \le 1$.
Thus, $f^\prime(0)$ must be non-negative, which means that
Eq.~(\ref{eq:q-less-divergence-contracting}) holds for $n=1$.

Next, we show that  (\ref{q-condition2}) implies (\ref{q-condition1}).
For any $P_U$ and $\{ \rho_A^u\}$, we have
\begin{eqnarray*}
I(U;B) &=& \sum_u P_U(u) D({\cal N}_B(\rho_A^u)\| {\cal N}_B(\bar{\rho}_A)), \\
I(U;E) &=& \sum_u P_U(u) D({\cal N}_E(\rho_A^u)\| {\cal N}_E(\bar{\rho}_A)),
\end{eqnarray*}
where 
\begin{eqnarray*}
\bar{\rho}_A = \sum_u P_U(u) \rho_A^u.
\end{eqnarray*}
Since Eq.~(\ref{eq:q-less-divergence-contracting}) holds for $n=1$, 
we have
\begin{eqnarray*}
I(U;B) \ge I(U;E).
\end{eqnarray*}
\end{proof}

\subsection{Proof of Theorem \ref{theorem:more-capable}}

It is a straight forward consequence of 
Proposition \ref{proposition:q-more-capable-no-preprocessing}.
Since ${\cal N}_B$ is more capable, Eq.~(\ref{eq:q-more-capable-alternative})
holds for every $n \ge 1$. Thus, we have
$C_p^{(1)}({\cal N}^{\otimes n}) = Q^{(1)}({\cal N}_B^{\otimes n})$
for every $n \ge 1$, and the statement of the theorem follows from 
Eqs.~(\ref{eq:private-capacity-formula}) and 
(\ref{eq:quantum-capacity-formula}). \qed

\subsection{Proof of Theorem \ref{theorem:less-noisy}}

Since less noisy implies more capable, by 
Theorem \ref{theorem:more-capable},
it suffice to show $Q({\cal N}_B) = Q^{(1)}({\cal N}_B)$.
For this purpose, we will show that
\begin{eqnarray}
\max_{\rho_{A^n}} I_c(A^n \rangle B^n) \le n \max_{\rho_A} I_c(A \rangle 
 B)
\label{eq:goal-main-theorem}
\end{eqnarray}
holds for every $n \ge 1$. For any input state
$\rho_{A^k A^\ell}$ on ${\cal H}_A^{\otimes (k+\ell)}$, let $\rho_{A^k}$ and $\rho_{A^\ell}$ be
the partial traces. Then, we have
\begin{eqnarray*}
\lefteqn{
I_c(A^k \rangle B^k) + I_c(A^\ell \rangle B^\ell) - I_c(A^k A^\ell 
\rangle B^k B^\ell) 
} \\
&=& D({\cal N}_B^{\otimes (k+\ell)}(\rho_{A^k A^\ell})\| {\cal N}_B^{\otimes 
 k}(\rho_{A^k}) \otimes {\cal N}_B^{\otimes \ell}(\rho_{A^\ell})) \\
&& - D({\cal N}_E^{\otimes (k+\ell)}(\rho_{A^k A^\ell})\| {\cal N}_E^{\otimes 
 k}(\rho_{A^k}) \otimes {\cal N}_E^{\otimes \ell}(\rho_{A^\ell})).
\end{eqnarray*}
Since Eq.~(\ref{eq:q-less-noisy-alternative}) holds for $n = (k+\ell)$, by ($n$ times extension of)
Proposition \ref{proposition:q-equivalence-less-noisy-divergence},
Eq.~(\ref{eq:q-less-divergence-contracting}) also holds for $n = (k+\ell)$,
which implies
\begin{eqnarray*}
I_c(A^k A^\ell \rangle B^k B^\ell) \le I_c(A^k \rangle B^k) + I_c(A^\ell 
 \rangle B^\ell).
\end{eqnarray*}
Thus, Eq.~(\ref{eq:goal-main-theorem}) can be proved by induction.
\qed

\section{Conclusion}
\label{section:conclusion}

In this paper, we introduced two new classes of
quantum channels, which we call more capable and
less noisy. For the more capable class, we showed
that the private capacity and quantum capacity coincide.
For the less noisy class, we showed that the private 
capacity and quantum capacity can be single letter 
characterized. 

Our results shed light on
further understanding of the private and quantum 
capacities of quantum channels. However, the conditions
such that a certain channel belongs to the more capable class or
the less noisy class are hard to verify in 
general, 
and we do not yet know whether there exists a channel that
belongs to the less noisy class but not to the degradable or
conjugate degradable classes, which is an important future
research agenda.

\section*{Acknowledgment}

This research is partly supported by
Grand-in-Aid for Young Scientists(B):2376033700
and Grant-in-Aid for Scientific Research(A):2324607101.
The author also would like to thank an anonymous reviewer
for helpful comments.


\begin{thebibliography}{24}
\expandafter\ifx\csname natexlab\endcsname\relax\def\natexlab#1{#1}\fi
\expandafter\ifx\csname bibnamefont\endcsname\relax
  \def\bibnamefont#1{#1}\fi
\expandafter\ifx\csname bibfnamefont\endcsname\relax
  \def\bibfnamefont#1{#1}\fi
\expandafter\ifx\csname citenamefont\endcsname\relax
  \def\citenamefont#1{#1}\fi
\expandafter\ifx\csname url\endcsname\relax
  \def\url#1{\texttt{#1}}\fi
\expandafter\ifx\csname urlprefix\endcsname\relax\def\urlprefix{URL }\fi
\providecommand{\bibinfo}[2]{#2}
\providecommand{\eprint}[2][]{\url{#2}}

\bibitem[{\citenamefont{Wyner}(1975)}]{wyner:75}
\bibinfo{author}{\bibfnamefont{A.~D.} \bibnamefont{Wyner}},
  \bibinfo{journal}{Bell Syst.~Tech.~J.} \textbf{\bibinfo{volume}{54}},
  \bibinfo{pages}{1355} (\bibinfo{year}{1975}).

\bibitem[{\citenamefont{Csisz\'ar and K\"orner}(1979)}]{csiszar:78}
\bibinfo{author}{\bibfnamefont{I.}~\bibnamefont{Csisz\'ar}} \bibnamefont{and}
  \bibinfo{author}{\bibfnamefont{J.}~\bibnamefont{K\"orner}},
  \bibinfo{journal}{IEEE Trans. Inform. Theory} \textbf{\bibinfo{volume}{24}},
  \bibinfo{pages}{339} (\bibinfo{year}{1979}).

\bibitem[{\citenamefont{Hayashi}(2006)}]{hayashi-book:06}
\bibinfo{author}{\bibfnamefont{M.}~\bibnamefont{Hayashi}},
  \emph{\bibinfo{title}{Quantum Information: An Introduction}}
  (\bibinfo{publisher}{Springer}, \bibinfo{year}{2006}).

\bibitem[{\citenamefont{Devetak}(2005)}]{devetak:05}
\bibinfo{author}{\bibfnamefont{I.}~\bibnamefont{Devetak}},
  \bibinfo{journal}{IEEE Trans. Inform. Theory} \textbf{\bibinfo{volume}{51}},
  \bibinfo{pages}{44} (\bibinfo{year}{2005}),
  \bibinfo{note}{arXiv:quant-ph/0304127}.

\bibitem[{\citenamefont{Cai et~al.}(2004)\citenamefont{Cai, Winter, and
  Yeung}}]{cai:04}
\bibinfo{author}{\bibfnamefont{N.}~\bibnamefont{Cai}},
  \bibinfo{author}{\bibfnamefont{A.}~\bibnamefont{Winter}}, \bibnamefont{and}
  \bibinfo{author}{\bibfnamefont{R.~W.} \bibnamefont{Yeung}},
  \bibinfo{journal}{Problems of Information Transmission}
  \textbf{\bibinfo{volume}{40}}, \bibinfo{pages}{26} (\bibinfo{year}{2004}).

\bibitem[{\citenamefont{Lloyd}(1997)}]{lloyd:97}
\bibinfo{author}{\bibfnamefont{S.}~\bibnamefont{Lloyd}},
  \bibinfo{journal}{Phys. Rev. A} \textbf{\bibinfo{volume}{55}},
  \bibinfo{pages}{1613} (\bibinfo{year}{1997}),
  \bibinfo{note}{arXive:quant-ph/9604015}.

\bibitem[{\citenamefont{Shor}(2002)}]{shor:02}
\bibinfo{author}{\bibfnamefont{P.~W.} \bibnamefont{Shor}}, in
  \emph{\bibinfo{booktitle}{Lecture Notes, MRSI Workshop on Quantum
  Computation}} (\bibinfo{year}{2002}).

\bibitem[{\citenamefont{Smith et~al.}(2008)\citenamefont{Smith, Renes, and
  Smolin}}]{smith:08}
\bibinfo{author}{\bibfnamefont{G.}~\bibnamefont{Smith}},
  \bibinfo{author}{\bibfnamefont{J.~M.} \bibnamefont{Renes}}, \bibnamefont{and}
  \bibinfo{author}{\bibfnamefont{J.~A.} \bibnamefont{Smolin}},
  \bibinfo{journal}{Phys. Rev. Lett.} \textbf{\bibinfo{volume}{100}},
  \bibinfo{pages}{170502} (\bibinfo{year}{2008}),
  \bibinfo{note}{arXiv:quant-ph/0607018}.

\bibitem[{\citenamefont{Shor and Smolin}(1996)}]{shor:96}
\bibinfo{author}{\bibfnamefont{P.~W.} \bibnamefont{Shor}} \bibnamefont{and}
  \bibinfo{author}{\bibfnamefont{J.~A.} \bibnamefont{Smolin}}
  (\bibinfo{year}{1996}), \bibinfo{note}{arXiv:quant-ph/9604006}.

\bibitem[{\citenamefont{DiVincenzo et~al.}(1998)\citenamefont{DiVincenzo, Shor,
  and Smolin}}]{divincenzo:98}
\bibinfo{author}{\bibfnamefont{D.~P.} \bibnamefont{DiVincenzo}},
  \bibinfo{author}{\bibfnamefont{P.~W.} \bibnamefont{Shor}}, \bibnamefont{and}
  \bibinfo{author}{\bibfnamefont{J.~A.} \bibnamefont{Smolin}},
  \bibinfo{journal}{Phys. Rev. A} \textbf{\bibinfo{volume}{57}},
  \bibinfo{pages}{830} (\bibinfo{year}{1998}),
  \bibinfo{note}{arXiv:quant-ph/9706061}.

\bibitem[{\citenamefont{Smith and Smolin}(2007)}]{smith:07b}
\bibinfo{author}{\bibfnamefont{G.}~\bibnamefont{Smith}} \bibnamefont{and}
  \bibinfo{author}{\bibfnamefont{J.~A.} \bibnamefont{Smolin}},
  \bibinfo{journal}{Phys. Rev. Lett.} \textbf{\bibinfo{volume}{98}},
  \bibinfo{pages}{030501} (\bibinfo{year}{2007}),
  \bibinfo{note}{arXiv:quant-ph/0604107}.

\bibitem[{\citenamefont{Devetak and Shor}(2005)}]{devetak:05b}
\bibinfo{author}{\bibfnamefont{I.}~\bibnamefont{Devetak}} \bibnamefont{and}
  \bibinfo{author}{\bibfnamefont{P.~W.} \bibnamefont{Shor}},
  \bibinfo{journal}{Comm. Math. Phys.} \textbf{\bibinfo{volume}{256}},
  \bibinfo{pages}{287} (\bibinfo{year}{2005}),
  \bibinfo{note}{arXiv:quant-ph/0311131}.

\bibitem[{\citenamefont{Br\'adler et~al.}(2010)\citenamefont{Br\'adler, Dutil,
  Hayden, and Muhammad}}]{bradler:10}
\bibinfo{author}{\bibfnamefont{K.}~\bibnamefont{Br\'adler}},
  \bibinfo{author}{\bibfnamefont{N.}~\bibnamefont{Dutil}},
  \bibinfo{author}{\bibfnamefont{P.}~\bibnamefont{Hayden}}, \bibnamefont{and}
  \bibinfo{author}{\bibfnamefont{A.}~\bibnamefont{Muhammad}},
  \bibinfo{journal}{Journal of Mathematical Physics}
  \textbf{\bibinfo{volume}{51}}, \bibinfo{pages}{072201}
  (\bibinfo{year}{2010}), \bibinfo{note}{arXiv:0909.3297}.

\bibitem[{\citenamefont{Schumacher and Westmoreland}(1998)}]{schumacher:98}
\bibinfo{author}{\bibfnamefont{B.}~\bibnamefont{Schumacher}} \bibnamefont{and}
  \bibinfo{author}{\bibfnamefont{M.~D.} \bibnamefont{Westmoreland}},
  \bibinfo{journal}{Phys. Rev. Lett.} \textbf{\bibinfo{volume}{80}},
  \bibinfo{pages}{5695} (\bibinfo{year}{1998}),
  \bibinfo{note}{arXiv:quant-ph/9709058}.

\bibitem[{\citenamefont{Shor and Preskill}(2000)}]{shor:00}
\bibinfo{author}{\bibfnamefont{P.~W.} \bibnamefont{Shor}} \bibnamefont{and}
  \bibinfo{author}{\bibfnamefont{J.}~\bibnamefont{Preskill}},
  \bibinfo{journal}{Phys. Rev. Lett.} \textbf{\bibinfo{volume}{85}},
  \bibinfo{pages}{441} (\bibinfo{year}{2000}),
  \bibinfo{note}{arXiv:quant-ph/0003004}.

\bibitem[{\citenamefont{Schumacher}(1996)}]{schumacher:96}
\bibinfo{author}{\bibfnamefont{B.}~\bibnamefont{Schumacher}},
  \bibinfo{journal}{Phys. Rev. A} \textbf{\bibinfo{volume}{54}},
  \bibinfo{pages}{2614} (\bibinfo{year}{1996}),
  \bibinfo{note}{arXiv:quant-ph/9604023}.

\bibitem[{\citenamefont{Smith}(2008)}]{smith:07}
\bibinfo{author}{\bibfnamefont{G.}~\bibnamefont{Smith}},
  \bibinfo{journal}{Phys. Rev. A} \textbf{\bibinfo{volume}{78}},
  \bibinfo{pages}{022306} (\bibinfo{year}{2008}),
  \bibinfo{note}{arXiv:0705.3838v1}.

\bibitem[{\citenamefont{Horodecki et~al.}(2008)\citenamefont{Horodecki,
  Pankowski, Horodecki, and Horodecki}}]{horodecki:08}
\bibinfo{author}{\bibfnamefont{K.}~\bibnamefont{Horodecki}},
  \bibinfo{author}{\bibfnamefont{L.}~\bibnamefont{Pankowski}},
  \bibinfo{author}{\bibfnamefont{M.}~\bibnamefont{Horodecki}},
  \bibnamefont{and}
  \bibinfo{author}{\bibfnamefont{P.}~\bibnamefont{Horodecki}},
  \bibinfo{journal}{IEEE Trans. Inform. Theory} \textbf{\bibinfo{volume}{54}},
  \bibinfo{pages}{2621} (\bibinfo{year}{2008}),
  \bibinfo{note}{arXiv:quant-ph/0506203}.

\bibitem[{\citenamefont{K\"orner and Marton}(1975)}]{korner:75}
\bibinfo{author}{\bibfnamefont{J.}~\bibnamefont{K\"orner}} \bibnamefont{and}
  \bibinfo{author}{\bibfnamefont{K.}~\bibnamefont{Marton}},
  \bibinfo{journal}{Keszthely Colloquium on Information Theory} pp.
  \bibinfo{pages}{411--423} (\bibinfo{year}{1975}).

\bibitem[{\citenamefont{Nielsen and Chuang}(2000)}]{nielsen-chuang:00}
\bibinfo{author}{\bibfnamefont{M.~A.} \bibnamefont{Nielsen}} \bibnamefont{and}
  \bibinfo{author}{\bibfnamefont{I.~L.} \bibnamefont{Chuang}},
  \emph{\bibinfo{title}{Quantum Computation and Quantum Information}}
  (\bibinfo{publisher}{Cambridge University Press}, \bibinfo{year}{2000}).

\bibitem[{\citenamefont{Cover and Thomas}(2006)}]{cover}
\bibinfo{author}{\bibfnamefont{T.~M.} \bibnamefont{Cover}} \bibnamefont{and}
  \bibinfo{author}{\bibfnamefont{J.~A.} \bibnamefont{Thomas}},
  \emph{\bibinfo{title}{Elements of Information Theory}}
  (\bibinfo{publisher}{John Wiley \& Sons}, \bibinfo{year}{2006}),
  \bibinfo{edition}{2nd} ed.

\bibitem[{\citenamefont{Horodecki et~al.}(2000)\citenamefont{Horodecki,
  Horodecki, and Horodecki}}]{horodecki:00}
\bibinfo{author}{\bibfnamefont{P.}~\bibnamefont{Horodecki}},
  \bibinfo{author}{\bibfnamefont{M.}~\bibnamefont{Horodecki}},
  \bibnamefont{and}
  \bibinfo{author}{\bibfnamefont{R.}~\bibnamefont{Horodecki}},
  \bibinfo{journal}{Journal of Modern Optics} \textbf{\bibinfo{volume}{47}},
  \bibinfo{pages}{347} (\bibinfo{year}{2000}),
  \bibinfo{note}{arXiv:quant-ph/9904092}.

\bibitem[{\citenamefont{van Dijk}(1997)}]{dijk:97}
\bibinfo{author}{\bibfnamefont{M.}~\bibnamefont{van Dijk}},
  \bibinfo{journal}{IEEE Trans. Inform. Theory} \textbf{\bibinfo{volume}{43}},
  \bibinfo{pages}{712} (\bibinfo{year}{1997}).

\bibitem[{\citenamefont{Ozel and Ulukus}(2011)}]{ozel:11}
\bibinfo{author}{\bibfnamefont{O.}~\bibnamefont{Ozel}} \bibnamefont{and}
  \bibinfo{author}{\bibfnamefont{S.}~\bibnamefont{Ulukus}}, in
  \emph{\bibinfo{booktitle}{Proc. IEEE Int. Symp. Inf. Theory 2011}}
  (\bibinfo{address}{Saintpeterburg, Russia}, \bibinfo{year}{2011}), pp.
  \bibinfo{pages}{627--631}, \bibinfo{note}{arXiv:1110.4613}.

\end{thebibliography}

\end{document}